\documentstyle[12pt]{article}

\input epsf.sty
\newcommand{\insertfig}[2]{\mbox{\epsfxsize=#1cm \epsfbox{#2.eps}}}
\def\1{\hbox{{1}\kern-.25em\hbox{l}}}

\begin{document}

\begin{titlepage}

\begin{flushright}
DOE/ER/40762-250 \\ [-2mm]
UMD-PP\#02-034 \\
\end{flushright}

\centerline{\large \bf The large-$N_c$ nuclear potential puzzle}

\vspace{15mm}

\centerline{\bf A.V. Belitsky, T.D. Cohen}

\vspace{5mm}

\centerline{\it Department of Physics}
\centerline{\it University of Maryland at College Park}
\centerline{\it College Park, MD 20742-4111, USA}

\vspace{15mm}

\centerline{\bf Abstract}

\vspace{1cm}

An analysis of the baryon-baryon potential from the point of view of large-$N_c$
QCD is performed. A comparison is made between the $N_c$-scaling behavior
directly obtained from an analysis at the quark-gluon level to the $N_c$-scaling
of the potential for a generic hadronic field theory in which it arises via meson
exchanges and for which the parameters of the theory are given by their canonical
large-$N_c$ scaling behavior. The purpose of this comparison is to use large-$N_c$
consistency to test the widespread view that the interaction between nuclei arises
from QCD through the exchange of mesons. Although at the one- and two-meson exchange
level the scaling rules for the potential derived from the hadronic theory matches
the quark-gluon level prediction, at the three- and higher-meson exchange level a
generic hadronic theory yields a potential which scales with $N_c$ faster than that
of the quark-gluon theory.

\vspace{40mm}


\vspace{5mm}


\end{titlepage}

%%%%%%%%%%%%%%%%%%%%%%%%%%%%%%%%%%%%%%%%%%%%%%%%%%%%%%%%%%%%%%%%%%%%%
\section{Hadronic dynamics and large-$N_c$ counting}
%%%%%%%%%%%%%%%%%%%%%%%%%%%%%%%%%%%%%%%%%%%%%%%%%%%%%%%%%%%%%%%%%%%%%

The absence of a calculational scheme for properties and low-energy
interactions of hadrons from the first principles of the underlying
microscopic theory of color dynamics calls for the treatment of this
domain by means of effective theories with degrees of freedom other
than quarks and gluons. Traditionally nuclear physicists have
envisioned nucleon-nucleon interactions as emerging from the exchange
of mesons. When QCD was established as the theory of strong interactions
three decades ago, the meson-exchange picture was viewed as arising
from it: QCD gives rise to effective hadronic degrees of freedom and
the interaction of these could then account for nuclear forces.
Obviously, these effective degrees of freedom can only describe the
low-lying modes of the theory where the underlying quark and gluon
degrees of freedom are not easy to disentangle. Therefore, one faces
the longstanding problem of correspondence between hadron and color
dynamics. The problem we wish to address is the extent to which QCD
justifies the traditional meson-exchange picture of nuclear forces.

The large-$N_c$ approximation provides a possible framework to investigate
this issue since there are a number of important simplifications of
QCD in this regime \cite{Hoo74,Wit79}. Of course, it is by no means
obvious that one can directly deduce specific phenomenological
consequences for the real world from the large-$N_c$ perspective.
Recall, for example, that the deuteron binding energy, $\varepsilon_B$,
is of order $N_c$ while the delta-nucleon mass splitting, $\delta M$, is
of order $1/N_c$. Thus, if one was in the large-$N_c$ world one would
expect a hierarchy of scales with $\varepsilon_B \sim {\cal O} (N_c)
\gg \delta M \sim {\cal O} (1/N_c)$. In the real world, however, one
finds these scales indeed widely separated but with the opposite order:
$\varepsilon_B \ll \delta M$. However, the argument that a QCD
description of low energy nucleon-nucleon interactions should be
describable in a meson-exchange picture does not appear to depend in
any explicit way on the fact that $N_c = 3$ in the physical world. Thus,
if the argument is valid it ought to apply equally well in a fictitious
multicolor world and one would then expect quantities deduced from the
meson-exchange picture to scale with $N_c$ in the same way as the same
quantities deduced from an analysis conducted directly at the quark-gluon
level. The use of large-$N_c$ scaling rules to test ideas from nuclear
physics is not new. It was argued more than a decade ago that nucleon
loop contributions based on point-like nucleon-meson couplings, as was
conventionally calculated in various quantum hadrodynamical models
\cite{SerWal86}, did not scale with $N_c$ in a manner consistent with
large-$N_c$ QCD and hence presumably did not capture the underlying QCD
dynamics \cite{Coh89}.

The physical spectrum of QCD consists of colorless hadronic states ---
baryons and mesons. As discussed by 't Hooft \cite{Hoo74} and Witten
\cite{Wit79}, large-$N_c$ QCD gives definite predictions for the scaling of
their characteristics with $N_c$. For example, the baryon and meson masses
are of order $N_c$ and unity, respectively, and the single-meson-baryon
coupling, $g_{1m}$, is of order $\sqrt{N_c}$ while meson-baryon scattering
amplitudes are of order unity. More  generally, reasoning along the lines
suggested by Witten \cite{Wit79} implies that the coupling of $N$ mesons to
a baryon scales at most as $g_{Nm} \sim N_c^{1 - N/2}$.

Apart from these generic counting rules there are additional constraints
coming from the spin-flavor structure of the interaction. This can be seen by
imposing the consistency of two single meson-baryon interactions vertices
$\mbox{\boldmath$V$}$ (nominally of order $N_c$ in total) with the meson-baryon
scattering which unitarity restricts to be of order unity. The
cancellations required for this to come about imply a contracted $SU(4)$
symmetry for two-flavor QCD \cite{GerSak83,GerSak84,DasMan93}, for a review,
see Refs.\ \cite{Man98,Jen98}. This implies that baryons form towers of
nearly degenerate states with $I = J$ and with splittings of order $1/N_c$.
The contracted $SU(4)$ relations hold for states in this tower. The
commutation algebra of the spin $\mbox{\boldmath$J$}$, isospin
$\mbox{\boldmath$I$}$ and spin-isospin $\mbox{\boldmath$X$}$ vertices is
given schematically by
\begin{equation}
[\mbox{\boldmath$I$}, \mbox{\boldmath$I$}]
\sim
\mbox{\boldmath$I$}
\, , \quad
[\mbox{\boldmath$J$}, \mbox{\boldmath$J$}]
\sim
\mbox{\boldmath$J$}
\, , \quad
[\mbox{\boldmath$I$}, \mbox{\boldmath$X$}]
\sim
\mbox{\boldmath$X$}
\, , \quad
[\mbox{\boldmath$J$}, \mbox{\boldmath$X$}]
\sim
\mbox{\boldmath$X$}
\, , \quad
[\mbox{\boldmath$X$}, \mbox{\boldmath$X$}]
\sim 1/N_c^2
\, ,
\end{equation}
The last relation implies a suppression when commutators arise. This occurs
in treatments of the tree-level baryon-meson scattering: for the
$N$-meson-baryon scattering amplitude ${\cal A}_N$ the destructive
interference \cite{ArnMat90,LamLiu96,LamLiu97} leads to the appearance of
multiple commutators of meson vertices leading to the consistency with the
large-$N_c$ counting rules predicting ${\cal A}_N \sim N_c^{1 - N/2}$.

\begin{figure}[t]
\begin{center}
\mbox{
\begin{picture}(0,70)(80,0)
\put(0,0){\insertfig{5}{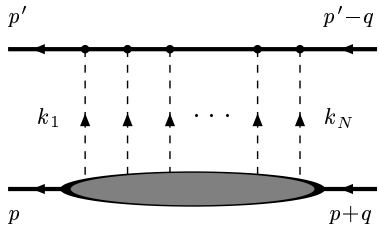}}
\end{picture}
}
\end{center}
\caption{\label{Generic} A generic exchange diagram for the baryon-baryon
scattering.}
\end{figure}

In the present study we address the issue of consistency of the large-$N_c$
QCD with the conventional meson-exchange picture used to describe nuclear
potentials. If the latter adequately describes the real world it must
possess the same multicolor asymptotics as deduced from QCD for loop
amplitudes. We consider the potential used for the baryon-baryon
scattering shown Fig.\ \ref{Generic}.  The problem at the quark level was
discussed by Kaplan and Savage in Ref.\ \cite{KapSav95} and by Kaplan and
Manohar in \cite{KapMan96}; collectively we refer to their analysis as KSM.
The basic strategy used by KSM was based on Witten's Hartree picture where
the interaction is identified as being due to the quark-line connected
diagrams. KSM then equate the nucleon-nucleon potential to the sum of
quark-line connected Feynman graphs which involve exchanges between two
groups of $N_c$ quark lines which represent baryons.  This is then analyzed
using the contracted $SU(4)$ symmetry. The principal results of this analysis
is that the strength of the spin-isospin structures of the nucleon-nucleon
potential scale as follows:
\begin{equation}
\label{KSM}
{\cal V}_{I = J} \sim N_c
\, , \qquad
{\cal V}_{I \neq J} \sim N_c^{-1}
\, ,
\end{equation}
where the subscript indicates the quantum numbers of the exchange in the
$t$-channel. It is straightforward to see from the large-$N_c$ scaling rules
of meson-baryon couplings that a one-meson exchange potential will satisfy
Eq.\ (\ref{KSM}). It is not immediately obvious, however, that multi-meson
exchange potentials will obey this rule since superficially they are clearly
larger than allowed by Eq.\ (\ref{KSM}). For example, at the two-meson
exchange level, both the retardation effects from box graphs and the
contributions from cross-box graphs enter ${\cal V}_{I = J}$ and
${\cal V}_{I \neq J}$ at order $N_c^2$. However, as shown in a detailed
calculation in Ref.\ \cite{BanCohGel01} cancellations between these two
yield potentials compatible with Eq.\ (\ref{KSM}). The contracted $SU(4)$
structure played an essential role in achieving this goal. Note that to
get the consistency with ${\cal V}_{I \ne J}$ cancellations up to order
$N_c^{- 3}$ are needed; in fact they occurred up to order $N_c^{- 4}$.

The results of Ref.\ \cite{BanCohGel01} raise the hope that similar
cancellations might be expected for all multi-meson exchanges. If these
were true, it would show consistency at large $N_c$ between the scaling of
the potential at the quark-gluon and hadronic level and would help to
justify meson-exchange based potential models as arising from QCD. However,
as will be shown in this paper, the type of cancellations seen for the
two-meson exchange, do not occur for general multi-meson exchanges. Thus,
potentials derived from generic hadronic theories calculated at a fixed
number of meson-exchanges do not give rise to potentials that respect the
KSM scaling rules of Eq.\ (\ref{KSM}).  This result is puzzling in view of
the general expectation that the physics of QCD at low energies can be
described in terms of hadronic degrees of freedom. We will refer to this
as the ``large-$N_c$ nuclear potential puzzle''.  There is a second puzzling
aspects of this problem. In studying exchanges of quark-antiquark pairs the
role of ``static'' pairs, i.e., pairs whose energy transfer is small (of order
$1/N_c$), is special: the leading order contribution of multi-pair exchanges
to the potential requires all pairs to be static. If there was a one-to-one
mapping between classes of quark-gluon diagrams with hadronic ones this would
correspond to static meson exchanges. In fact, however we will see that the
``dangerous'' contributions at the hadronic level --- the ones which contradict
the KSM rules of Eq.\ (\ref{KSM}) --- all come from non-static meson exchanges.

The calculations of multi-meson exchanges can get quite complicated.
Accordingly, it is useful to develop tools which greatly simplify the
analysis. The non-abelian generalization of the eikonal formula
\cite{LamLiu97} which can be used to compute the sum of certain
multi-meson amplitudes will be quite handy in this context. In the
following section we will review this formalism. In section \ref{Scattering}
we show how sums of various multi-meson diagrams lead to contributions which
are incompatible with the KSM rules. Finally, we conclude with a discussion
of possible resolutions of this apparent paradox.

%%%%%%%%%%%%%%%%%%%%%%%%%%%%%%%%%%%%%%%%%%%%%%%%%%%%%%%%%%%%%%%%%%%%%
%            Figure 1
%%%%%%%%%%%%%%%%%%%%%%%%%%%%%%%%%%%%%%%%%%%%%%%%%%%%%%%%%%%%%%%%%%%%%
%%%%%%%%%%%%%%%%%%%%%%%%%%%%%%%%%%%%%%%%%%%%%%%%%%%%%%%%%%%%%%%%%%%%%

%%%%%%%%%%%%%%%%%%%%%%%%%%%%%%%%%%%%%%%%%%%%%%%%%%%%%%%%%%%%%%%%%%%%%
\section{Non-abelian eikonal formula}
\label{EikonalFormula}
%%%%%%%%%%%%%%%%%%%%%%%%%%%%%%%%%%%%%%%%%%%%%%%%%%%%%%%%%%%%%%%%%%%%%

Following Ref.\ \cite{LamLiu97} we consider a baryon which emits or
absorbs a number of virtual mesons. Since the baryon is extremely heavy
at $N_c = \infty$ it can be treated as almost static. It will be a
nonrelativistic particle if its three-momentum $\mbox{\boldmath$p$}$ is
of ${\cal O} (N_c^0)$; the four momentum is then $(M + \mbox{\boldmath$p$}^2
/ \left( 2 M \right), \mbox{\boldmath$p$})$. In this kinematic regime the
heavy baryon propagator is approximated by the product of the conventional
eikonal propagator and a projection matrix (which will be omitted later) on
the large components of the nucleon bispinor
\begin{equation}
\label{EikonalPropagator}
\frac{1}{ {\not\!p} + {\not\!k} - M^\prime + i \epsilon}
\to
\frac{1}{\omega + i \epsilon} \frac{\1 + \gamma_0}{2}
\, .
\end{equation}
Taken literally, this expression which neglects the nucleon recoil is only
valid for meson energies $\omega$, $k = (\omega, \mbox{\boldmath$k$})$, of order
${\cal O} (N_c^0)$. If $\omega$ is of order $N_c^{-1}$ then the denominator
of the propagator (\ref{EikonalPropagator}) is modified to include the recoil
effect as follows: $\omega \to \omega + \delta M + \left( \mbox{\boldmath$p$}^2
- (\mbox{\boldmath$p$} + \mbox{\boldmath$k$})^2 \right) / \left( 2 M \right)$,
with the mass difference of the ``degenerate'' states in the baryon tower
$\delta M = M - M^\prime$. Note, $\delta M$ is ${\cal O} (N_c^{-1})$
\cite{Jen93,CarGeoOso93,LutMar93}. For simplicity we generally omit this
modification in our expressions. This approximation has to be kept in mind
since it leads to singularities in the integrand of certain loop amplitudes.
However, they are easily identifiable and cured by the simple expedient of
reintroducing the recoil correction.

There is an important kinematical constraint in this regime. Since the initial
and final nucleons are on-shell and have three-momenta of order $N_c^0$, the
kinetic energies of the initial and final states are thus of order $N_c^{-1}$
(due to the fact that $M \sim N_c$) and thus the total exchenged energy is
also of order $N_c^{-1}$.

The blob on the lower baryon line in Fig.\ \ref{Generic} represents the
tree amplitude for production of $N$ mesons ${\cal A}_N$; it includes
all possible permutation of single meson emissions from the baryon line and
this contribution reads
\begin{equation}
\label{N-meson-A}
{\cal A}_N \equiv \sum_{\sigma
\in \{ 1 \dots N \}} {\cal A} [\sigma_1 \cdots \sigma_N]
\, ,
\end{equation}
where
\begin{equation}
{\cal A} [\sigma_1 \cdots \sigma_N]
=
\mbox{\boldmath$V$}_{\!\!\sigma_1} \mbox{\boldmath$V$}_{\!\!\sigma_2}
\dots
\mbox{\boldmath$V$}_{\!\!\sigma_N}
a [\sigma_1 \cdots \sigma_N]
\, ,
\end{equation}
with one of the matrices $\mbox{\boldmath$V$} = \left\{ \mbox{\boldmath$1$} ,
\ \mbox{\boldmath$I$} , \ \mbox{\boldmath$J$} , \ \mbox{\boldmath$X$}
\right\}$ associated with the meson-baryon interaction vertex, and
\begin{equation}
a [\sigma_1 \cdots \sigma_N]
\equiv
- 2 \pi i \ \delta \left( \sum_{j = 1}^N \omega_j \right)
\prod_{j = 1}^{N - 1}
\frac{1}{\sum_{k = 1}^{j} \omega_{\sigma_k} + i \epsilon}
\, .
\end{equation}
Obviously, for a single element $\sigma$ (and neglecting recoil as discussed
above), $a [\sigma] = - 2 \pi i \delta (\omega_\sigma)$.

Due to the destructive Bose-Einstein interference, ${\cal A}_N$ can be
expressed by a multiple commutator formula for the sum of non-abelian eikonal
amplitudes discussed in Ref.\ \cite{LamLiu96},
\begin{eqnarray}
{\cal A}_N = \sum_{\sigma \in \{ 1 \dots N \}}
\widetilde {\cal A} [\sigma_1 \cdots \sigma_N]
\, .
\end{eqnarray}
The constructive definition of the multi-commutator amplitude $\widetilde
{\cal A}$ goes as follows. For a given ordering of lines, if the rightmost
element of the given permutation is smaller than any other element to its
left, then there is only one partition and it is the whole tree. Otherwise,
go to the first element whose number is smaller and draw a partition just
before that element. Next, start to the left of this partition and repeat
the procedure again. Through the cut separating the partitions, the amplitude
factorizes:
\begin{equation}
\widetilde {\cal A} [ \sigma_1 \cdots \sigma_2 | \sigma_3 \cdots
\sigma_4 | \cdots ]
=
\widetilde {\cal A} [ \sigma_1 \cdots \sigma_2 ]
\widetilde {\cal A} [ \sigma_3 \cdots \sigma_4 ]
\widetilde {\cal A} [ \cdots ]
\, ,
\end{equation}
where $\sigma_2 < \sigma_4 < \cdots$, $\sigma_1 > \cdots > \sigma_2$ and
$\sigma_3 > \cdots > \sigma_4$. The amplitudes without partitions are
given by the commutator formula with the innermost commutator formed by
the last two elements of the tree,
\begin{equation}
\widetilde {\cal A} [ \sigma_1 \cdots \sigma_{n - 1} \sigma_n ]
=
[ \mbox{\boldmath$V$}_{\!\!\sigma_1} ,
\cdots ,
[ \mbox{\boldmath$V$}_{\!\!\sigma_{n - 1}}, \mbox{\boldmath$V$}_{\!\!\sigma_n} ]
\cdots ]
a [\sigma_1 \cdots \sigma_{n - 1} \sigma_n]
\, .
\end{equation}
The first two nontrivial examples read
\begin{eqnarray}
{\cal A}_2
\!\!\!&=&\!\!\!
\widetilde {\cal A} [1 | 2]
+
\widetilde {\cal A} [21]
\, , \\
{\cal A}_3
\!\!\!&=&\!\!\!
\widetilde {\cal A} [1 | 2 | 3]
+
\widetilde {\cal A} [21 | 3]
+
\widetilde {\cal A} [1 | 32]
+
\widetilde {\cal A} [231]
+
\widetilde {\cal A} [31 | 2]
+
\widetilde {\cal A} [321]
\, . \nonumber
\end{eqnarray}
The proof is straightforward by the Fourier transformation of both sides
of the equality.

The generalizations are obvious. The constructive formula reads
\begin{equation}
\label{NonAbelianEikonal}
{\cal A}_N = \sum_{\sigma = \sigma_1 + \dots + \sigma_j}
\left( \prod_{k = 1}^{j} a [\{ \sigma_k \}] \right)
\mbox{\boldmath$V$} \left[ \{ \sigma_1 \} \right]
\cdots
\mbox{\boldmath$V$} \left[ \{ \sigma_j \} \right]
\, ,
\end{equation}
with a number of partitions $j$ of a given permutation $\sigma$ into
sets $\sigma_k$ with elements $\{ \sigma_k \} = \sigma_k^{(1)} , \cdots ,
\sigma_k^{(k)}$ constructed according to the definition given above.
For a single element in the set $\mbox{\boldmath$V$}\left[ k \right]$
coincides with  the single vertex $\mbox{\boldmath$V$}_k$. For more
than one element it is given by a multiple commutator $\mbox{\boldmath$V$}
\left[ \{ \sigma_k \} \right] = [ \mbox{\boldmath$V$}_{\!\!\sigma_{k}^{(1)}} ,
\cdots , [ \mbox{\boldmath$V$}_{\!\!\sigma_k^{(k - 1)}},
\mbox{\boldmath$V$}_{\!\!\sigma_k^{(k)}} ] \cdots ]$. For the abelian case,
i.e., when $\mbox{\boldmath$V$} \to \mbox{\boldmath$1$}$, all commutators
vanish and one gets the well-known eikonal formula.

%%%%%%%%%%%%%%%%%%%%%%%%%%%%%%%%%%%%%%%%%%%%%%%%%%%%%%%%%%%%%%%%%%%%%
\section{Baryon-baryon scattering}
\label{Scattering}
%%%%%%%%%%%%%%%%%%%%%%%%%%%%%%%%%%%%%%%%%%%%%%%%%%%%%%%%%%%%%%%%%%%%%

%%%%%%%%%%%%%%%%%%%%%%%%%%%%%%%%%%%%%%%%%%%%%%%%%%%%%%%%%%%%%%%%%%%%%
%            Figure 2
%%%%%%%%%%%%%%%%%%%%%%%%%%%%%%%%%%%%%%%%%%%%%%%%%%%%%%%%%%%%%%%%%%%%%
\begin{figure}[t]
\begin{center}
\mbox{
\begin{picture}(0,95)(190,0)
\put(0,0){\insertfig{13}{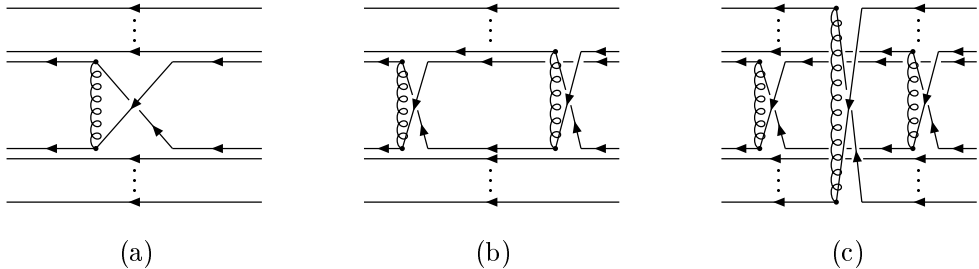}}
\end{picture}
}
\end{center}
\caption{\label{Quark-Exchange} Quark diagrams which can be interpreted
as meson exchanges.}
\end{figure}
%%%%%%%%%%%%%%%%%%%%%%%%%%%%%%%%%%%%%%%%%%%%%%%%%%%%%%%%%%%%%%%%%%%%%

The baryon-baryon interaction at large $N_c$ comes from the exchange of
pairs of quark constituents. A quark from one of the hadrons switches
places with any quark in the other baryon and exchanges a gluon to
neutralize the color charge. The quark exchange, in Fig.\
\ref{Quark-Exchange} (a), may naturally be reinterpreted at the
hadronic level as a meson exchange,
\begin{equation}
{\cal V} (\mbox{\boldmath$q$})
= g_{1m}^2
D ( - \mbox{\boldmath$q$}^2 )
\,
\quad\mbox{with}\quad
D (q^2) = \frac{1}{q^2 - m^2 + i \epsilon}
\, ,
\end{equation}
where for simplicity we have chosen the example of the scalar-isoscalar
channel. Clearly, this correspondence between the meson and quark exchanges
is not valid on a diagram-by-diagram basis.  Presumably the sum of all
quark exchange diagrams including an arbitrary number of gluons connecting
the exchanged quarks (which if planar are leading order in $N_c$) gets
mapped onto the sum of all one meson-exchange graphs for mesons with these
quantum numbers.

In analyzing the contributions to the potential one must recall that
the full amplitude is obtained from the potential by an iteration. Thus,
nonrelativistic amplitudes satisfy the Lippmann-Schwinger equation
\begin{equation}
{\cal T} ( \mbox{\boldmath$p$}, \mbox{\boldmath$p$} + \mbox{\boldmath$q$})
= -
{\cal V}
( \mbox{\boldmath$q$} )
+
\int \frac{d^3 \mbox{\boldmath$k$}}{(2 \pi)^3}
{\cal V}
( \mbox{\boldmath$k$} )
{\cal G}
( \mbox{\boldmath$k$} )
{\cal T}
(
\mbox{\boldmath$p$} + \mbox{\boldmath$k$},
\mbox{\boldmath$p$} + \mbox{\boldmath$q$}
)
\, ,
\end{equation}
with the potential ${\cal V}$ and the two-baryon propagator
\begin{equation}
\label{Two-Baryon-Prop}
{\cal G} ( \mbox{\boldmath$k$} )
\equiv
\frac{
1
}{
\frac{
\mbox{\boldmath$\scriptstyle p$}^2
}{
M
}
-
\frac{
( \mbox{\boldmath$\scriptstyle k$} + \mbox{\boldmath$\scriptstyle p$} )^2
}{
M
}
+
i \epsilon
}
=
\int \frac{d \omega}{2 \pi i}
\frac{1}{
\left(
\omega
-
\frac{\mbox{\boldmath$\scriptstyle p$}^2}{2 M}
+
\frac{( \mbox{\boldmath$\scriptstyle k$} + \mbox{\boldmath$\scriptstyle p$} )^2}{2 M}
- i \epsilon
\right)
\left(
\omega
+
\frac{\mbox{\boldmath$\scriptstyle p$}^2}{2 M}
-
\frac{( \mbox{\boldmath$\scriptstyle k$} + \mbox{\boldmath$\scriptstyle p$} )^2}{2 M}
+ i \epsilon
\right)}
\, .
\end{equation}
Here we defined ${\cal G}$ as the conventional nonrelativistic Green function
for a two-baryon system of the reduced mass $M/2$ and in the last form we
reexpressed it in terms of the single-particle propagators neglecting $\delta M$.
The kinematics is chosen according to Fig.\ \ref{Generic} where $p = (E,
\mbox{\boldmath$p$})$ and $p^\prime = (E, - \mbox{\boldmath$p$})$ with
nonrelativistic expansion $E \approx M + \mbox{\boldmath$p$}^2/ \left( 2 M \right)$.
The momentum transferred $q = p - p^\prime = (q_0, \mbox{\boldmath$q$})$ has a
small time-component $q_0 \sim {\cal O} (1/N_c)$ since it is inversely proportional
to the baryon mass.

When assessing the contribution of some hadronic-level Feynman diagram to
the potential, it is essential to note that the Feynman diagrams sum to
give the full amplitude and not just the potential. Accordingly, to extract
the contributions to the potential, one must remove all contributions which
correspond to iterates of the potential. Fortunately, they are easily
identifiable. From the nonrelativistic form (\ref{Two-Baryon-Prop}) it is
clear that as $N_c \rightarrow \infty$, ${\cal G}$ diverges since $M \sim N_c$.
Thus, in the large-$N_c$ limit, these potential iterates are associated
with infrared singularities and these are the only infrared singularities in
the problem. Thus, when one encounters them in the integrals one may make
the substitution
\begin{equation}
\label{G}
\int d \omega \frac{\delta (\omega)}{\omega + i \epsilon}
\rightarrow
{\cal G}
\, ,
\end{equation}
where one inputs ${\cal G}$ with appropriate kinematics. More significantly,
having identified these terms as arising from a potential iterate, one can
remove them from the amplitude to extract an irreducible contribution to the
potential only.

\begin{figure}[t]
\begin{center}
\mbox{
\begin{picture}(0,50)(190,0)
\put(0,0){\insertfig{13}{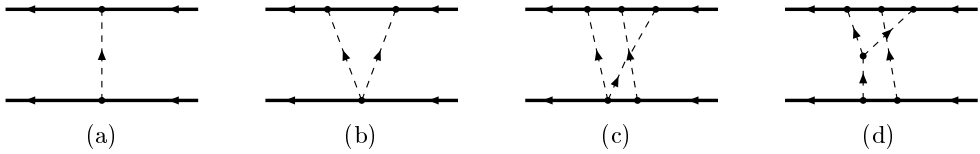}}
\end{picture}
}
\end{center}
\caption{\label{Three-Meson} Typical meson-exchange diagrams contributing
to the baryon-baryon scattering amplitude. They are expected to correspond
to the quark-pair exchange graphs in Fig.\ \ref{Quark-Exchange}. The last
two graphs (c) and (d) violate the KSM large-$N_c$ counting rule.}
\end{figure}

The baryon-baryon interaction at the QCD level is generated by various
complicated quark-exchange processes sampled in Fig.\ \ref{Quark-Exchange}.
One expects that these are connected diagrams which contribute to the
potential while the quark-line disconnected amplitudes, such as
\ref{Quark-Exchange} (c), are associated with iterates of the potential
which arise in the Lippmann-Schwinger equation. On the hadronic side, the
sum of the quark-level diagrams can presumably be interpreted as meson
exchanges. A few examples of these can be found in Fig.\ \ref{Three-Meson}.
As we will demonstrate momentarily, at multiple meson exchanges the
correspondence between quark and hadron level diagrams is lost.

%%%%%%%%%%%%%%%%%%%%%%%%%%%%%%%%%%%%%%%%%%%%%%%%%%%%%%%%%%%%%%%%%%%%%
%            Figure 3
%%%%%%%%%%%%%%%%%%%%%%%%%%%%%%%%%%%%%%%%%%%%%%%%%%%%%%%%%%%%%%%%%%%%%
%%%%%%%%%%%%%%%%%%%%%%%%%%%%%%%%%%%%%%%%%%%%%%%%%%%%%%%%%%%%%%%%%%%%%

For the sake of concreteness, we illustrate the issues by considering
bosons with scalar-isoscalar quantum numbers. For other channels, the
presence of nontrivial spin-isospin indices introduces significant
complications due to the noncommutativity of vertices but does not
affect the main line of reasoning. We will return to the effects of
noncommutativity when we consider ladder and crossed-ladder diagrams.
We also will neglect all effects of momentum dependence in the
meson-baryon couplings. Again it easy to see that they do not alter
our conclusions.

Let us analyze what kinds of quark configurations give contributions to
the potential. The one quark-pair exchange in Fig.\ \ref{Quark-Exchange}
(a) is translated into the one-meson exchange of Fig.\ \ref{Three-Meson}
(a). At the quark level, there is a $N_c^2$ combinatorial factor from the
number of possibilities to join the quark lines in both baryons and there
is a $1/\left( \sqrt{N_c} \right)^2$ from the two gluon couplings. This
results in an overall scaling as $N_c$ which is compatible with Eq.\
(\ref{KSM}). On the hadronic side the meson-baryon coupling behaves as
$\sqrt{N_c}$ and the one-meson exchange diagram thus also scales as $N_c$.
Note that the energy transfer scales as $1/N_c$ and vanishes for $N_c \to
\infty$.  Thus, both the exchanged meson and the exchanged quark pair are
static with this kinematics.

The two quark-pair exchange diagram in Fig.\ \ref{Quark-Exchange} (b) is
of order $N_c$: a combinatoric factor of $N_c^3$ and a factor of
$1/\left( \sqrt{N_c} \right)^4$ from the coupling constants. Note, this
diagram has both pairs of quarks coupled to the same quark line in one of
the baryons and thus corresponds to the diagram with a two-meson-baryon
(seagull) vertex, Fig.\ \ref{Three-Meson} (b). The reason for both pairs
to couple to the same line in a baryon is simple. If they do not, then the
graph would be quark-line disconnected and thus, by hypothesis, not
associated with the irreducible part of the potential. By the $N_c$-counting
for this disconnected quark graph to agree with the seagull diagram, it
would require an extra gluon exchange to yield the same total large-$N_c$
behavior. On the quark level one again realizes that the leading order
contribution of this type comes from the exchange of static pairs. In this
case the requirement goes beyond simple kinematics. Kinematically the energies
of the two exchanges must be equal and opposite (up to $1/N_c$-corrections)
but need not be static. However, the two affected quark lines in the upper
baryon in Fig.\ \ref{Quark-Exchange} (b) do not communicate by a gluon and,
therefore, if each quark exchange carries an energy then the energy of the
final state quarks would not correspond to the single particle energies in
the Hartree ground state. On the other hand, if they do exchange a gluon
there is an extra $1/N_c$-suppression without an off-setting combinatoric gain
so the term is not of leading order in the $1/N_c$-expansion. On the hadronic
level the static nature of the exchange can be seen to arises as follows from
the Feynman diagram of Fig.\ \ref{Three-Meson} (b):
\begin{eqnarray}
i {\cal T}
\!\!\!&=&\!\!\!
g_{1m}^2 g_{2m}^{}
\int
\prod_{j = 1}^2 \frac{d^3 \mbox{\boldmath$k$}_j}{(2 \pi)^3}
(2 \pi)^3
\delta^{(3)}
\left(
\mbox{\boldmath$k$}_1 + \mbox{\boldmath$k$}_2 - \mbox{\boldmath$q$}
\right)
\int
\prod_{j = 1}^2 \frac{d \omega_j}{2 \pi}
(2 \pi)
\frac{\delta ( \omega_1 + \omega_2 )}{\omega_1 - i \epsilon}
D (k_1^2) D (k_2^2)
\\
&=&\!\!\!
i \pi g_{1m}^2 g_{2m}^{}
\int
\prod_{j = 1}^2 \frac{d^3 \mbox{\boldmath$k$}_j}{(2 \pi)^3}
(2 \pi)^3
\delta^{(3)}
\left(
\mbox{\boldmath$k$}_1 + \mbox{\boldmath$k$}_2 - \mbox{\boldmath$q$}
\right)
\int
\prod_{j = 1}^2 \frac{d \omega_j}{2 \pi}
(2 \pi) \delta (\omega_1) \delta (\omega_2)
D (- \mbox{\boldmath$k$}_1^2)
D (- \mbox{\boldmath$k$}_2^2)
\, . \nonumber
\end{eqnarray}
where $g_{1m} \sim \sqrt{N_c}$ is the single meson-nucleon coupling
constant, $g_{2m} \sim N_c^0$ is the ``seagull'' coupling and $D(-
\mbox{\boldmath$k$}^2) = \left (- \mbox{\boldmath$k$}^2 - m^2 + i \epsilon
\right )^{-1}$ is the meson propagator for a meson of mass $m$,
three-momentum $\mbox{\boldmath$k$}$ and zero energy.  The second equality
follows from the fact that the meson propagators are even functions
of the meson energies $\omega_j$ so that one immediately finds that the
principal value part of the baryon propagator cancels under integration
and only the $\delta$-function piece survives. Thus only static-exchange
mesons contribute.

In addition to the seagull-type two-meson exchange there are also box and
crossed-box contributions to the potential. As discussed at length in Ref.\
\cite{BanCohGel01}, the retardation effects in the box and crossed-box
graphs separately violate the counting rules of Eq.\ (\ref{KSM}).
Moreover the contributions from these graphs come entirely from non-static
meson exchanges. However, these terms when summed cancel one another
yielding the total result consistent with Eq.\ (\ref{KSM}) and without
contributions from non-static mesons. At the level of the amplitude at the
two-meson exchange level one also has an iterate of the static one-meson
exchanges. This term presumably is associated with the non-quark-line-connected
part of the two quark-pair exchange graphs at the quark-gluon level.

%%%%%%%%%%%%%%%%%%%%%%%%%%%%%%%%%%%%%%%%%%%%%%%%%%%%%%%%%%%%%%%%%%%%%
\subsection{A problem at three-meson exchange level}
\label{MMM-Problem}
%%%%%%%%%%%%%%%%%%%%%%%%%%%%%%%%%%%%%%%%%%%%%%%%%%%%%%%%%%%%%%%%%%%%%

The three quark-pair exchange of the type displayed in Fig.\
\ref{Quark-Exchange} (c) is not quark-line connected.  Rather two of the
exchanges are connected while the third one is disconnected. By hypothesis,
then, this graph is not to be associated with the potential but presumably
with a potential iterate. Using the analysis similar to what was done
previously in this paper it is easy to see that all of the exchanged
quark-pairs must be static at leading order in the $1/N_c$-expansion (which
for this class of graph is formally of order $N_c^2$ --- larger than the
allowed scaling of the potential --- due to the disconnected nature of the
graph.)

%%%%%%%%%%%%%%%%%%%%%%%%%%%%%%%%%%%%%%%%%%%%%%%%%%%%%%%%%%%%%%%%%%%%%
%            Figure 4
%%%%%%%%%%%%%%%%%%%%%%%%%%%%%%%%%%%%%%%%%%%%%%%%%%%%%%%%%%%%%%%%%%%%%
\begin{figure}[t]
\begin{center}
\mbox{
\begin{picture}(0,140)(160,0)
\put(0,0){\insertfig{11}{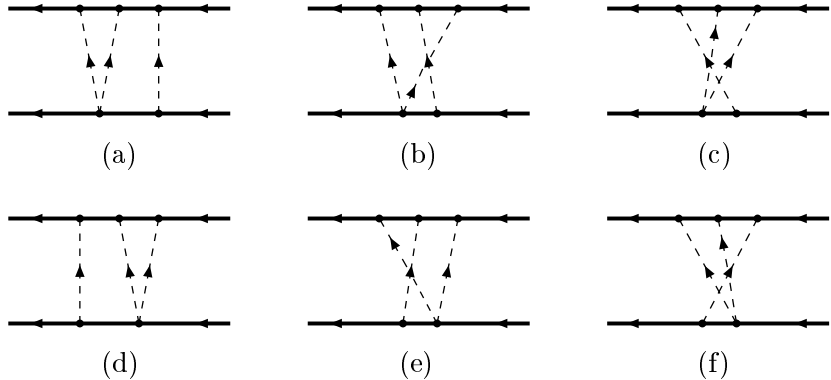}}
\end{picture}
}
\end{center}
\caption{\label{Seagull} Three-meson exchange diagrams with one seagull
vertex.}
\end{figure}
%%%%%%%%%%%%%%%%%%%%%%%%%%%%%%%%%%%%%%%%%%%%%%%%%%%%%%%%%%%%%%%%%%%%%

At the hadronic level this graph should be associated with a seagull exchange
as in Fig.\ \ref{Quark-Exchange} (b) dressed with an extra single-meson exchange.
This generates  six diagrams displayed in Fig.\ \ref{Seagull}. The generalized
eikonal formula of section \ref{EikonalFormula} can be applied to describe the
lower line in Fig.\ \ref{Seagull} in a straightforward manner (with the
``seagull'' vertex taking the place of a single-particle one)\footnote{We use
the eikonal formula on the lower line since it is simpler. When applied to the
upper line, the potential iterate contributions are not identifiable as easily.}.
Because the vertices commute, only the $\delta$-function contributes for the
propagators in appropriate pairs of graphs. This implies a cancellation of the
fully non-static parts of diagrams Fig.\ \ref{Seagull} (a) with \ref{Seagull} (c)
and \ref{Seagull} (d) with \ref{Seagull} (f), where ``fully non-static'' implies
that all the mesons exchanged are non-static. These cancellations are reminiscent
of the ones of the retardation effects in box graphs against crossed-box graphs
discussed in Ref.\ \cite{BanCohGel01}. The $\delta$-function contributions
to Fig.\ \ref{Seagull} (a,c) and Fig.\ \ref{Seagull} (d,f) give combinations of
propagators of the form seen in Eq.\ (\ref{G}). Obviously, they are iterates of
the lowest order Lippmann-Schwinger kernel and do not affect the potential itself.
Finally, there is a $\delta$-function contribution to Fig.\ (\ref{Seagull}) (b,e).
This term is not an iterate of the potential but rather an irreducible contribution
to it given by
\begin{eqnarray}
\delta {\cal V} \!\!\!&=&\!\!\! - g_{1m}^4 g_{2m}^{}
\int
\prod_{j = 1}^{3} \frac{d^3 \mbox{\boldmath$k$}_j}{(2 \pi)^3}
\, (2 \pi)^3 \, \delta^{(3)}
\left( \sum_{j = 1}^{3}
\mbox{\boldmath$k$}_j - \mbox{\boldmath$q$} \right)
\nonumber\\
&\times&\!\!\!
\int \prod_{j = 1}^{3} \frac{d \omega_j}{2 \pi} \,
(2 \pi) \delta \left( \sum_{j = 1}^{3} \omega_j \right )
\left( \prod_{j = 1}^3 D (k_j^2) \right) \,
\frac{1}{\omega_1 - i \epsilon}
\frac{1}{\omega_1 + \omega_2 - i \epsilon} (- 2 \pi i) \delta (\omega_2)
\, .
\end{eqnarray}
Note, there are two $\delta$-functions in energy but three energies.
Thus, they do not force all three mesons to be static.  Rather, the single
exchanged meson is static, while the two mesons connected to the seagull
vertex have equal and opposite (but generally non-zero) energy transfer.
Performing the energy integrations yields
\begin{equation}
\label{deltaV}
\delta {\cal V}
=
-  g_{1m}^4 g_{2m}^{}
\int
\prod_{j = 1}^{3} \frac{d^3 \mbox{\boldmath$k$}_j}{(2 \pi)^3}
\, (2 \pi)^3 \, \delta^{(3)}
\left( \sum_{j = 1}^{3}
\mbox{\boldmath$k$}_j - \mbox{\boldmath$q$} \right)
\, D ( - \mbox{\boldmath$k$}_2^2 )
{\cal R}_2 (\varepsilon_1 , \varepsilon_3)
\end{equation}
For simplicity, have introduced a general function ${\cal R}_n$
which shows commonly in expressions for multi-meson contributions with
two non-static mesons,
\begin{equation}
{\cal R}_n (\varepsilon_j , \varepsilon_k)
\equiv \int \frac{d \omega}{2 \pi i}
\frac{1}{\left( \omega - i \epsilon \right)^n}
D (\omega^2 - \mbox{\boldmath$k$}_j^2)
D (\omega^2 - \mbox{\boldmath$k$}_k^2)
= \frac{1}{2 \left( \varepsilon_j \varepsilon_k \right)^{n + 1}}
\frac{
\varepsilon_j^{n + 1} -  \varepsilon_k^{n + 1}
}{
\varepsilon_j^2 -  \varepsilon_k^2
}
\, ,
\end{equation}
with $\varepsilon_j \equiv \sqrt{\mbox{\boldmath$k$}_j^2 + m^2}$.

A couple comments is in order at this point. The first is that $\delta {\cal V}$
as given in Eq.\ (\ref{deltaV}) scales as $N_c^2$. This is easily seen since $g_{1m}
\sim \sqrt{N_c}$ while $g_{2m} \sim N_c^0$ and both $D ( - \mbox{\boldmath$k$}_2^2 )$
and ${\cal R}_n (\varepsilon_j , \varepsilon_k)$ are independent of $N_c$. The
second is that this contribution to $\delta {\cal V}$ does not vanish. Both
$D ( - \mbox{\boldmath$k$}_2^2 )$ and ${\cal R}_n (\varepsilon_j , \varepsilon_k)$
are positive definite so no cancellations are possible in the integral.
This is the heart of our puzzle. The three-meson exchange contribution to the
potential is of order $N_c^2$ this is incompatible with Eq.\ (\ref{KSM}) which
was derived at the quark level. Moreover, the contribution comes entirely from
non-static meson contributions. At the three quark-pair exchange level the
amplitude only gets contributions from non-static pairs at order $N_c$ (even
including quark-line disconnected pieces.)

In the explicit computation done above, point-like meson-nucleon couplings
were used.  It is clear that had momentum dependent vertices been included
the functional form of the integral used to derive the potential would be
altered but the $N_c$-counting would not. Similarly, it is clear that this
problem is not restricted to effects from seagull graphs. It is trivial to see
that order $N_c^2$ contributions will arise in other topologies with
multi-meson couplings such as exemplified in Fig.\ \ref{Three-Meson} (d).
Note, though, that this diagram corresponds to the same quark-level topology
in Fig.\ \ref{Quark-Exchange} (c) hinting on a possible conspiracy of the
seagull and three-meson vertex contributions.

%%%%%%%%%%%%%%%%%%%%%%%%%%%%%%%%%%%%%%%%%%%%%%%%%%%%%%%%%%%%%%%%%%%%%
\subsection{Ladders and crossed ladders}
%%%%%%%%%%%%%%%%%%%%%%%%%%%%%%%%%%%%%%%%%%%%%%%%%%%%%%%%%%%%%%%%%%%%%

A similar problem shows up in the contributions form ladder and crossed-ladder
diagrams. Here, we will consider the effect of $N$ identical mesons with
non-derivative point-like couplings to the baryons. Derivative couplings
will not alter the conclusions, but, as seen in Ref.\ \cite{BanCohGel01},
greatly complicate the analysis. Non-point-like couplings can also be easily
included and do not alter the qualitative results either. Similarly, the
restriction to identical mesons is done for simplicity; again the
conclusions will not be strongly dependent on this. Since we are interested
in the possibility of effects which violate the KSM rules by having a
``super-leading'' $N_c$-dependence, we will consider the exchanges of mesons
which have couplings to baryons of order $\sqrt{N_c}$, the maximum allowable.
From the analysis of the contracted $SU(4)$ symmetry in Ref.\ \cite{DasJenMan93}
these will be a scalar-isoscalar vertex or a spin one, isospin one
$\mbox{\boldmath$X$}_{i a}$-type. As will be seen below, the super-leading
effect depends on the non-vanishing of the commutators of the vertices.
Thus, the scalar-isoscalar exchanges, which commute, will not contribute
(unless other non-commuting exchanges are also present). Accordingly we
will restrict our attention to mesons that couple in a non-derivative way
to $\mbox{\boldmath$X$}_{i a}$. An example of such a meson is the spatial
part of the ${\sf A}_1$ (recalling that for nonrelativistic baryons, the
couplings to the temporal and spatial parts of vector mesons can be separated).

Note that our result will depend on the fact that our problem is
$[\mbox{\boldmath$X$}_{i a}, \mbox{\boldmath$X$}_{j b}] \neq 0$. Of course,
for contracted $SU(4)$ this commutator is zero. However, one only has the
contracted $SU(4)$ symmetry for infinite $N_c$. For finite $N_c$ the
commutator is small, scaling as $N_c^{-2}$ but not zero. If the commutator
is multiplied by a function which grows with $N_c$ rapidly enough it will
contribute and can indeed be associated with super-leading effects.

\begin{figure}[t]
\begin{center}
\mbox{
\begin{picture}(0,50)(150,0)
\put(0,0){\insertfig{11}{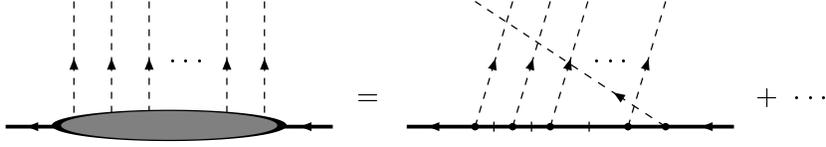}}
\end{picture}
}
\end{center}
\caption{\label{Ladder} A crossed-ladder diagram (on the right-hand
side of the equality) which generates a super-leading contribution to
the baryon potential. Dashes on the baryon line denote the on-shell
condition (the cuts) of the corresponding propagators.}
\end{figure}

For any given number of meson exchanges  one has to sum the blob in Fig.\
(\ref{Ladder}) representing the emission of mesons with all orderings. For
this purpose the non-abelian generalization of the eikonal formula in
Eq.\ (\ref{NonAbelianEikonal}) is indispensable.

%%%%%%%%%%%%%%%%%%%%%%%%%%%%%%%%%%%%%%%%%%%%%%%%%%%%%%%%%%%%%%%%%%%%%
%            Figure 5
%%%%%%%%%%%%%%%%%%%%%%%%%%%%%%%%%%%%%%%%%%%%%%%%%%%%%%%%%%%%%%%%%%%%%
%%%%%%%%%%%%%%%%%%%%%%%%%%%%%%%%%%%%%%%%%%%%%%%%%%%%%%%%%%%%%%%%%%%%%

The amplitude for the exchange of $N$ point-coupled ${\sf A}_1$ mesons
is given by
\begin{eqnarray}
\label{lad}
{\cal T}_N \!\!\!&=&\!\!\! (- i)^N g_{1m}^{2 N}
\int \prod_{j = 1}^{N} \frac{d^3 \mbox{\boldmath$k$}_j}{(2 \pi)^3}
\, (2 \pi)^3 \, \delta^{(3)}
\left( \sum_{j = 1}^{N}
\mbox{\boldmath$k$}_j - \mbox{\boldmath$q$} \right)
\int \prod_{j = 1}^{N}
\frac{d \omega_j}{2 \pi}
\left( \prod_{j = 1}^N D (k_j^2) \right)
\nonumber\\
&\times&\!\!\!
\mbox{\boldmath$V$}_{\!\!1} \frac{1}{\omega_1 - i \epsilon}
\mbox{\boldmath$V$}_{\!\!2} \frac{1}{\omega_1 + \omega_2 - i \epsilon}
\mbox{\boldmath$V$}_{\!\!3}
\cdots
\mbox{\boldmath$V$}_{\!\!N - 1} \frac{1}{\sum_{j = 1}^{N - 1} \omega_j - i \epsilon}
\mbox{\boldmath$V$}_{\!\!N} \otimes {\cal A}_N
\, ,
\end{eqnarray}
where ${\cal A}_N$ is the sum of all permutations of the positions of
the mesons coupling to the lower line, see Eq.\ (\ref{N-meson-A}). The
generalization of the eikonal formula in Eq.\ (\ref{NonAbelianEikonal})
gives a straightforward way to express ${\cal A}_N$. The notation in
Eq.\ (\ref{lad}) is to be understood as follows. The vertex factors,
$\mbox{\boldmath$V$}_{\!\!N}$ represent the spin-isospin structure of
the couplings. Since, in this example we are considering ${\sf A}_1$
exchange, at leading order in $N_c$, it is simply represented by the
$\mbox{\boldmath$X$}_{i a}$ of the contracted $SU(4)$ symmetry. The tensor
product is used to describe the spin-isospin structures which appear on
each of two baryon lines. Note that ${\cal A}_N$ also contains the
structures $\mbox{\boldmath$V$}_{\!\!1} \cdots \mbox{\boldmath$V$}_{\!\!N}$.
There is an implicit contraction of spin and isospin indices when we
write $\mbox{\boldmath$V$}_{\!\! k} \otimes \mbox{\boldmath$V$}_{\!\!k}$,
so that for this example
\begin{eqnarray*}
\mbox{\boldmath$V$}_{k} \otimes \mbox{\boldmath$V$}_{k}
\equiv
\sum_{i a} \mbox{\boldmath$X$}_{i a} \otimes \mbox{\boldmath$X$}_{i a}
\, .
\end{eqnarray*}
This contraction comes about for obvious reasons: when an ${\sf A}_1$ meson
is exchanged the spatial $i$ and isospin $a$ components couple to both the
upper and lower baryon lines.

The generalized eikonal formula allows us to evaluate ${\cal A}_N$ is a
straightforward way. There is a large number of terms contained in
${\cal A}_N$.  These can be organized by the number of commutators they
contain.  From the form of Eq.\ (\ref{NonAbelianEikonal}) one sees that
each time a commutator is added there is one fewer $\delta$-function in
energy. The term with no commutators has a total of $N$ $\delta$-functions.
Thus, these meson exchanges are static. However, it is
straightforward to see that this term when combined with the meson
propagators and the upper baryon line contains $N - 1$ combinations
of propagators which are divergent in the infrared and have the form of
Eq.\ (\ref{G}). They obviously correspond to $N$ iterates of the one-meson
exchange potential. (This can be checked by adding back the recoil
corrections and observing that Lippmann-Schwinger propagator
emerges.) Thus, this term does not contribute to the potential. Next
there are terms with a single commutator in ${\cal A}_N$. The term with
the commutator $[\mbox{\boldmath$V$}_{\!\!i},\mbox{\boldmath$V$}_{\!\!j}]$
(where $i$ and $j$ label the position of the meson vertex in the standard
ordering, $i > j$) contains $N - 1 - (i - j)$ combinations of the form of
Eq.\ (\ref{G}). For $i - j < N - 1$, these terms again are iterates of some
lower meson-exchange potentials and do not contribute to the potential itself.
However, for $i - j = N - 1$, i.e., where $i = N$ and $j = 1$, there are
no such combinations of propagators and, thus, this does not correspond to
an iterate and directly contributes to the potential. This term is unique
and leads to a contribution of the form:
\begin{eqnarray}
\label{Vlad}
\delta {\cal V}_N
\!\!\!&=&\!\!\! (- 1)^{N + 1} g_{1m}^{2 N}
\int \prod_{j = 1}^{N} \frac{d^3 \mbox{\boldmath$k$}_j}{(2 \pi)^3}
\, (2 \pi)^3 \, \delta^{(3)}
\left( \sum_{j = 1}^{N}
\mbox{\boldmath$k$}_j - \mbox{\boldmath$q$} \right)
\nonumber\\
&\times&\!\!\!
\mbox{\boldmath$V$}_{\!\!1}
\cdots
\mbox{\boldmath$V$}_{\!\!N}
\otimes
[\mbox{\boldmath$V$}_{\!\!N}, \mbox{\boldmath$V$}_{\!\!1}]
\mbox{\boldmath$V$}_{\!\!2} \cdots \mbox{\boldmath$V$}_{\!\!N - 1}
\left( \prod_{j = 2}^{N - 1} D (- \mbox{\boldmath$k$}_j^2) \right)
{\cal R}_N ( \varepsilon_1, \varepsilon_N )
\, .
\end{eqnarray}
Note that apart from the $\mbox{\boldmath$V$}$-factors (which are
multiplicative) the integrand is positive definite. Thus $\delta {\cal V}$
cannot vanish after integration. Recall that the commutator of two
$\mbox{\boldmath$X$}$'s is of order $1/N_c^2$. Note also that there are
contractions of the spin and isospin on the lower baryon line with the
upper. Thus the commutator on the lower line  induces a commutator on
the upper line in a fashion similar to that seen in Ref.\ \cite{BanCohGel01}.
Thus one expects the commutators to give rise to an overall suppression
factor of $N_c^{-4}$. The coupling constants $g_{1m}$ scales as $\sqrt{N_c}$.
Combining the suppression due to the commutator with the coupling constants
produces $\delta {\cal V}_N \sim N_c^{N-4}$ in the potential. For $N > 3$ this
is incompatible with KSM scaling rules of Eq.\ (\ref{KSM}).

One could continue in the application of the non-abelian generalization
of the eikonal formula in the evaluation of ${\cal A}_N$. All additional
terms will have two or more  commutators. Some of these terms will
correspond to potential iterates, but some will be contributions to the
potential. Recall, however, that commutators typically lead to suppression
factors in the large-$N_c$ expansion. For example, two single commutators
will be suppressed from the single commutator by a factor of $N_c^{-2}$
(with an additional factor $N_c^{-2}$ induced on the upper line. Similarly
a triple commutator is suppressed by a factor of $N_c^{-2}$. Such
contributions are therefore subleading compared to the result in Eq.\
(\ref{Vlad}). However, the double commutator is not down by powers of
$N_c^{-1}$. It is a simple exercise to see that such term have a different
dependence on the momentum transfer than the contribution in Eq.\
(\ref{Vlad}). Thus, such a contribution  cannot generally cancel terms
coming from a single commutator. Thus one concludes that the sum of ladder
and crossed-ladder diagrams with $N$ rungs contributes to the potential
with the strength $\delta {\cal V}_N \sim N_c^{N - 4}$.

%%%%%%%%%%%%%%%%%%%%%%%%%%%%%%%%%%%%%%%%%%%%%%%%%%%%%%%%%%%%%%%%%%%%%
\section{Discussion}
%%%%%%%%%%%%%%%%%%%%%%%%%%%%%%%%%%%%%%%%%%%%%%%%%%%%%%%%%%%%%%%%%%%%%

As discussed in the introduction, the principal reason for undertaking
the present investigation is to try to understand whether the traditional
meson-exchange picture of nucleon-nucleon forces can be understood as
arising from QCD. The central idea was that the general argument that
a meson exchange picture ought to work was not based on the details of
QCD other than confinement and hence ought to work for all $N_c$. Thus,
the fact that the $N_c$-counting of the nucleon-nucleon potential
calculated at the quark-gluon level does not match the $N_c$-counting
based on a meson-exchange picture might be taken as a strong evidence
against the latter. The conclusion that QCD is not compatible with a
meson-exchange dynamics of nucleon-nucleon interactions at low momenta
is quite radical. In the first place it goes against the conventional
wisdom of nuclear physics. Moreover, it might be troubling from the
perspective of large-$N_c$ QCD. Here-to-fore in all known examples in the
purely mesonic sector the large-$N_c$ QCD counting matches what would be
found in a purely hadronic theory with parameters scaling in a manner
consistent with large-$N_c$ QCD. In the baryon-number one sector problems
discussed so far in the literature have a correspondence between hadronic
and quark-gluon based descriptions. This was demonstrated so far for the
tree-level Compton (multi-) meson-baryon scattering amplitudes
\cite{GerSak83,GerSak84,DasMan93,LamLiu97} and chiral corrections to decay
constants \cite{DasMan93}. Note, however, that on the basis of the
considerations advocated presently one expects a potential inconsistency
in the meson-baryon Compton amplitude from the loop diagrams of the type
\ref{Three-Meson} (d) (obviously, with the bottom baryon line being removed).
Before one accepts this radical conclusion, it is essential to explore other
ways to resolve the puzzle of why the two descriptions have $N_c$-scaling
behaviors which do not match. There are a number of possible explanations
which are consistent with what we know about the system. However, all of them
are unattractive in one way or another.

One general class of possibilities is that the way the hadronic-level
calculation is organized is in some way defective and this hides
cancellations which might bring consistency. Here we have seen that
diagrams for generic exchanges of mesons of some fixed type do not
cancel among themselves. It is certainly logically possible that they
may cancel with some other class of graphs to preserve the scaling results
of Eq.\ (\ref{KSM}). One might hope that there is some way to reorganize
the calculation so that the cancellations do occur. We see no simple way
for this to come about. We do see two obvious scenarios where these
cancellations could come about but have serious drawbacks as a resolution.

In the first scenario the cancellations would still occur for generic meson
coupling constants with the large-$N_c$ rules but would require a larger class
of graphs. Since additional meson exchanges lead to increasingly super-leading
terms it is conceivable that they can be resummed. Such a resummation could
lead to a small result. If for example, the series were essentially geometric
--- $N_c + N_c^2 + N_c^3 + ... $ --- one could sum it to $1/(1 - N_c) - 1$
which goes to (minus) one in the large-$N_c$ limit. Unfortunately, this scenario
has a manifest drawback: we see no reason from the mathematical structures as
to why we might expect this to happen.

A second scenario is that the cancellations will not occur for a generic
hadronic theory with parameters consistent with general $N_c$-counting
rules but depend rather on a conspiracy between the coupling constants,
masses and so on for the various mesons to yield cancellations. As a
matter of principle, of course, this cannot be ruled out. Two-flavor
isospin-symmetric QCD is a theory with essentially two free parameters
$\Lambda_{\rm QCD}$ and $m_q$ (defined with some scheme) and all hadronic
parameters derived from QCD depend on these two in a very complicated way.
Thus, there are very complex correlations between the hadronic parameters
and one could imagine that these correlations conspire to enforce massive
cancellations between diagrams with different topologies on the hadronic
level. Since we do not know the structure of this theory, it is very
difficult to conceive of how such cancellations would come about in practice
and why they would hold for arbitrary momentum transfers. We note also that
if this scenario was correct, it becomes hard to justify the use of the meson
exchange in practice. In any practical implementation of a potential based
on a meson exchange, the number of mesons included and the forms of the
particular interactions are necessarily restricted. It is very implausible
that the restricted form chosen would be capable of enforcing these nongeneric
cancellations.

An alternative class of explanations focuses on the validity of KSM rules. Eq.\
(\ref{KSM}) is supposed to apply in the kinematic regime $\mbox{\boldmath$p$}
\sim N_c^0$. However it is conceivable that this regime is simply not suitable
for a large-$N_c$ expansion. It was long ago noted by Witten that scattering
observables cannot have a smooth large-$N_c$ limit in this kinematic regime
\cite{Wit79}, thus Refs.\ \cite{KapSav95,KapMan96} focus on the potential rather
than the amplitude. However, there has never been a systematic demonstration
that the potential in this regime has a smooth limit. Of course, the derivation
of Eq.\ (\ref{KSM}) is very plausible. It is based on the Hartree picture which
in turn implies that only quark-line connected graphs contribute.  It is worth
noting, however, that despite its plausibility, the derivation may be flawed.
Witten justified the Hartree picture for $N_c$ quarks in their ground state
where it can be shown that non-Hartree type correlations are suppressed in the
$1/N_c$-expansion. On the other hand, the potential only has meaning as an
ingredient in a Schr\"odinger (or Lippmann-Schwinger) equation. The Schr\"odinger
wave function implies strong correlations between the nucleons which at the
quark level does not correspond to a Hartree-type wave function. Thus, the
question arises of whether the Hartree/quark-line connected approximation can
be justified. Indeed at a philosophical level one might ask whether the potential,
which, after all, is not experimentally accessible in any direct way, can even be
assigned an $N_c$-scaling. At a more practical level the problem is that
quark-line disconnected pieces certainly contribute to amplitudes but by
hypothesis do not contribute to the potential. Thus, for the approach to be
consistent, these contribution must be associated with iterates of the potential.
However, there is no general argument of which we aware of which demonstrates
that the quark-line disconnected contributions are in fact described by
iterating the potential.

Explanations of the mismatch in $N_c$ counting based on the possibility
that Eq.\ ({\ref{KSM}) is invalid face a major hurdle. The problem is that
most likely that Eq.\ (\ref{KSM}) could fail is that its derivation assumes
that the quark-line disconnected parts are associated with potential
iterates and that this assumption could be wrong. However, this mismatch
between the $N_c$-dependence at the quark-gluon level from the hadronic
one occurs even at the amplitude level where the question of determining
which contribution is an iterate does not arise. Note that although the
amplitude for the three quark-pair exchange is of order $N_c^3$, this
contribution comes from purely quark-line disconnected contributions and
for the reason discussed above is necessarily associated with the static
quark-pair exchanges. The leading contribution arising from two pairs
being non-static and one being static is of order $N_c$ (see Fig.\
\ref{Quark-Exchange} (c) and add a gluon connecting exchanged quarks in
the top baryon line in the subgraph (b) of (c)). However at the hadronic
level the term with a static and two non-static mesons, as discussed in
section \ref{MMM-Problem}, contributed to the total amplitude at order
$N_c^2$. Thus, even at the amplitude level the contribution from the
exchange of one static and two non-static mesons does not match the
contribution from the exchange of two non-static quark pairs and a
static pair. This mismatch cannot be ascribed to the distinction in how
things are apportioned between the potential and its iterates.

In summary, the mismatch between the $N_c$-counting of contributions to the
potential between quark- and hadron-based descriptions remains puzzling. It
seems likely to us that an understanding the roles played by static and
non-static exchanges of mesons in the hadronic picture and quark-antiquark
pairs in the quark-based formalism is essential resolving this puzzle
definitively. The resolution of the problem is of real importance as it
provides insights into the relationship of nuclear phenomenology to QCD.

\vspace{0.5cm}

We would like to thank X. Ji who initiated the present collaboration.
We acknowledge conversations with B.A. Gelman, D.B. Kaplan, A.V. Manohar
and S.J. Wallace.

\end{document}